\documentclass[
preprint,
jcp,
aip,
]{revtex4-1}

\usepackage{amssymb}
\usepackage{placeins}
\usepackage{bm}
\usepackage{amsmath}
\usepackage{color}
\usepackage{graphicx}
\usepackage{mathtools}
\usepackage{gensymb}
\usepackage[table]{xcolor}
\usepackage{array}
\usepackage{makecell}
\usepackage{sistyle}
\usepackage{amsfonts}
\usepackage{multirow}
\usepackage[utf8]{inputenc}
\usepackage[T1]{fontenc}
\usepackage{lmodern}
\usepackage[outdir=./,update]{epstopdf}
\usepackage[english]{babel}
\usepackage{caption}
\usepackage{subcaption}
\usepackage{bm}

\begin{document}


\title{Cold reactive and non-reactive collisions of Li and Rb with C$_{2}^{-}$: implications for hybrid trap experiments}

\author{Milaim Kas}
\email{milakas@ulb.ac.be}
\author{J\'er\^ome Loreau}
\author{Jacques Li\'evin}
\author{Nathalie Vaeck}
\affiliation{
 Service de Chimie Quantique et Photophysique (CQP)\\
 Universit\'e libre de Bruxelles (ULB), Brussels, Belgium 
}

\date{\today}

\begin{abstract}
We present a theoretical investigation of cold reactive and non-reactive collisions of Li and Rb atoms with C$_{2}^{-}$. 
The potential energy surfaces for the singlet and triplet states of the Li--C$_{2}^{-}$ and Rb--C$_{2}^{-}$ systems have been obtained using the CASSCF/ic-MRCI+Q approach with extended basis sets. 
The potential energy surfaces are then used to investigate the associative detachment reaction and to calculate rotationally inelastic cross sections at low collision energies by means of the close-coupling method. 
The results are compared to those obtained for other anionic systems such as Rb-OH$^{-}$, and the implications for hybrid trap experiments and sympathetic cooling experiments are explored. 
Furthermore, we discuss the possibility to perform Doppler thermometry on the C$_{2}^{-}$ anion and investigate the collision process involving excited electronic states. 
\end{abstract}

\maketitle


\section{\label{sec:int}Introduction}
The study of atomic and molecular dynamics in the cold and ultracold regime, which started several decades ago, is a very prolific domain of research. The ability to confine and cool down atoms at temperatures below $\mu$K temperatures using laser schemes has resulted in numerous exciting breakthroughs while the study of cold molecules, which present additional internal degrees of freedom, have led to interesting new areas of research \cite{Carr2009,Dulieu2011,Krems2010} ranging from precision spectroscopy \cite{Loh2013} and test of fundamental constants \cite{Bethlem2009} to quantum control of chemical reactions \cite{Quemener2012}. A broad range of experimental methods such as Stark deceleration \cite{VandeMeerakker2006}, sympathetic cooling \cite{Asvany2009,Zipkes2011}, and even direct laser cooling \cite{Kozyryev2017}, can be used to cool down molecular gases. In particular, molecular ions offer interesting possibilities since they can be easily trapped using radio frequency \cite{Wester2009a} or Penning \cite{Blaum2010} traps. This allows subsequent cooling using a cold (such as cryogenic helium) or ultracold (such as laser-cooled atoms) buffer gas. So far, most experimental groups have been focusing on molecular cations. However, following the discovery of anions in various astrochemical environments \cite{Chaizy1991,McCarthy2006,Vuitton2009}, the proposal of laser cooling schemes for atomic \cite{Walter2014} and molecular \cite{Yzombard2015} anions, and the prospects of sympathetic cooling anti-matter using cold anions \cite{Kellerbauer2009}, an increasing theoretical and experimental effort has been devoted to negatively charged systems. 
The most general approach to produce cold ions is sympathetic cooling. One of the most important feature is that radio frequency traps are state-independent and usually have rather deep wells \cite{Wester2009a} (around 1 eV), which prevents inelastic collisions from causing trap losses of the ions. However, reactive collisions may still lead to losses through charge transfer or other mechanisms that involve neutralisation of the ions. In the case of anions, one such mechanism is the associative electronic detachment (AED) reaction, AB$^{-}$ + C $\rightarrow$ ABC + $e^{-}$. The reaction occurs through the crossing of the ABC and ABC$^{-}$ potential energy surfaces (PES), defining the autodetachment (AD) region where the anionic state is embedded in the continuum of the neutral state and can spontaneously eject its excess electron \cite{Chen1967}. The autodetachment rate is very large when the AD region is reached\cite{Simons1998}: often around 10$^{14}-10^{15}$ s$^{-1}$. In order to get a low AED rate, one general rule would be to choose a closed shell AB$^{-}$ species with a large electron affinity. Several such candidates have been recently proposed for sympathetic cooling with ultracold alkali and alkaline earth atoms \cite{Tomza2017}. However, the position of the AD region will still depend on the atomic collision partner. For example, trap losses have been experimentally observed for the Rb + OH$^{-}$ system \cite{Deiglmayr2012}. We have previously shown that choosing another alkali or alkaline earth atomic species would not lead to detachment \cite{Kas2017} since the vertical detachment energy of MOH$^{-}$ depends on M. An other example is the CN$^{-}$ anion for which the AED reaction can occur for some collision partners despite having a large electron affinity (3.862 eV) \cite{Jerosimic2018,Khiri2015}. In conclusion, it is difficult to rely on general molecular properties to predict whether or not to expect AED reactions at low temperatures without explicitly calculating both the anion and neutral PES. This is especially true when dealing with open shell species. \\

In the present paper, we investigate the possibility to sympathetic cool the C$_{2}^{-}$ anion using ultracold Rb or Li atoms. The choice of Rb is motivated by its current use in hybrid trap experiments involving molecular anions \cite{Deiglmayr2012} while Li is the lightest alkali. The results for other alkali are expected to lie somewhere in between, as shown for collisions of OH$^-$ with alkali atoms \cite{Kas2017,Tomza2017}. The molecular anion C$_2^-$ has been widely studied both theoretically \cite{Shi2016a,Nichols1987,W.Thulstrup1974,Aventini1975} and experimentally \cite{Bragg2003,Jones1980,Endres2014,Barckholtz2001,Mead1985}, in particular in the context of astrochemistry since it has been predicted to exist (but not yet detected) in some astrophysical environments \cite{Civis2005}. C$_{2}^{-}$ is an open shell species with $^{2}\Sigma^{+}_{g}$ symmetry and an electron affinity of 3.2 eV \cite{Shi2016a}. It possesses several electronic excited states that could be used in laser cooling schemes \cite{Yzombard2015,Fesel2017,Gerber2018}. Furthermore, these excited states may be used to perform fluorescence imaging and Doppler spectroscopy in the context of sympathetic cooling experiments \cite{Gianfrani2016}. This non-destructive method should allow to map the position of the anions and estimate their translational temperature. It could be compared to destructive measurements such as photodetachment tomography \cite{Deiglmayr2012}. Thus, in addition to threshold photodetachment spectroscopy \cite{Endres2014}, several ways of measuring the translational and rotational temperature of C$_{2}^{-}$ can be used. Furthermore, the ability to control the excited state of Li or Rb through a tuning of the magneto-optical trap (MOT) laser's intensity or via the repumping laser when a dark spontaneous optical trap (dark SPOTs) scheme is used  \cite{Holtkemeier2017}, as well as C$_{2}^{-}$ using an electronic transition, opens up the possibility to study excited state dynamics.  \\  
In a ideal situation, the energy is thermally distributed along the different degrees of freedom and the ions should quickly thermalise with the buffer gas. However, in practice, the presence of the laser field, the coupling of the ions motion with the rf field \cite{Cetina2012,Holtkemeier2016}, collisions with the background gas \cite{Ravi2012}, reactive collision with the buffer gas, and black-body radiation inside the reactive chamber lead to a non-equilibrium situation. Therefore, the cross sections and rate constants related to the different processes are needed to correctly describe such an environment. \\

This work is structured as follows: in section \ref{PES}, we present the calculation of the potential energy surfaces for the Rb--C$_{2}^{-}$ and Li--C$_{2}^{-}$ complexes. In section \ref{Excited}, we investigate the AED reaction, both from ground state and excited state channels. In section \ref{Dyn}, we use the PESs to perform close-coupling calculations and extract rotationally inelastic cross sections and rate constants corresponding to the (de-)excitation of C$_{2}^{-}$ in collisions with Li and Rb atoms.

\section{\label{sec:Comp}Potential energy surfaces}
\label{PES}
\begin{center}
\begin{figure}[htbp!]
\includegraphics[bb=169 0 233 242,scale=0.4]{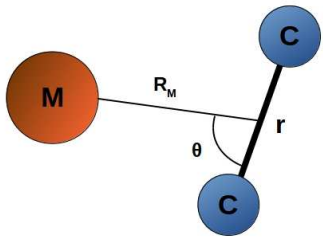}
\caption{Jacobi coordinates defining the M-C$_{2}^{-}$ molecular system, where M = Li or Rb.}
\label{coord}
\end{figure}
\end{center}

The interaction between Rb($^{2}S$) or Li($^{2}S$) and C$_{2}^{-}(^{2}\Sigma^{+}_{g})$ leads to a singlet and triplet $\Sigma^{+}$ molecular states for the linear case (C$_{\infty v}$ point group) which become $A'$ states at bent geometries (C$_{s}$ point group) and $A_{1}$ for the ``T shaped'' structure (C$_{2v}$ point group). All our calculations have been performed in the C$_{s}$ point group, and subsequent symmetry considerations will therefore be discussed using the C$_{s}$ terminology. The singlet and triplet molecular states arise from different electronic configurations. For the RbC$_{2}^{-}$ singlet state the valence electronic configuration is $8a'^{2}9a'^{2}2a''^{2}10a'^{0}$, where the first 3 molecular orbitals are mainly formed by the 2$p$ atomic orbitals of the carbon atoms. The main electronic configuration of the triplet molecular state is $8a'^{2}9a'^{1}2a''^{2}10a'^{1}$ where the 10$a'$ molecular orbital has a large contribution from the 5$s$ atomic orbital of Rb. Both configurations become degenerate at large distance and adiabatically lead to the C$_{2}^{-} (^{2}\Sigma^{+}_{g})+$Rb$(^{2}S)$ dissociation channel. The same is also true for the LiC$_{2}^{-}$ states. This system is a typical multireference case where single reference method such as the coupled cluster approach will not be suited. Therefore, the complete active space self consistent field (CASSCF) method has been used to obtain a set of state-averaged molecular orbitals. For RbC$_{2}^{-}$ the active space includes 10 $a'$ and 3 $a''$ molecular orbitals, corresponding to the valence atomic orbitals of the two C atoms and the $4s$, $4p$ and $5s$ atomic orbitals of Rb. For LiC$_{2}^{-}$, the active space includes all orbitals of Li and the valence orbitals of the C atoms, resulting in 8 $a'$ and 2 $a''$ molecular orbitals. The internally contracted multireference configuration interaction with single and double excitation (ic-MRCI) is then applied to the CASSCF wave function for each state \cite{Knowles1992}. In addition, a correction for the size inconsistency problem (Davidson correction) has been taken into account \cite{Langhoff1974,Meissner1988,Werner2008}. This correction is usually denoted by MRCI+Q since it allows to account for the effect of quadruple excitations. The AWCV5Z basis set \cite{ThomH.Dunning1988} has been used for the Li atoms with inclusion of the $1s$ electrons into the correlation treatment. The Rb atom is described by the ECPMDF28 (shortened MDF) electron core potential (ECP) \cite{Lim2005}. The latter allows the $4s$, $4p$ and $5s$ electrons to be explicitly correlated. The corresponding orbitals are described by the companion \textit{spdfg} basis set of the MDF ECP \cite{Lim2005}. Following our previous work on alkali and alkaline earth hydroxides \cite{Kas2017,Kas2016}, both Rb and Li basis sets have been extended with 3\textit{s},2\textit{p},1\textit{s}, and 1\textit{f} even-tempered diffuse functions. 
We have also investigated the influence of the core 1$s_{\textrm{C}}$ electron correlation on the PES, which is often neglected. When including the 1$s_{\textrm{C}}$ orbitals, we have used the core correlation consistent AWCVQZ basis set for the C atoms while the AVQZ has been used otherwise. 
All calculations have been performed using the \begin{small}MOLPRO\end{small} 2012 package \cite{MOLPRO}.
In order to compute the potential energy surface (PES) for further scattering calculations, the M-C$_{2}^{-}$ system (with M=Li or Rb) is described by the Jacobi coordinates $R_{\mathrm{M}}$, $\theta$, and $r$ illustrated in Figure \ref{coord}. Since we will investigate low temperature collisions, the vibration of C$_2^-$ is not expected to be important and the distance $r$ was held fixed at\cite{Mead1985} 1.268 $\angstrom$ so that the C$_{2}^{-}$ molecule will be treated as a rigid rotor. 

\begin{figure}[htbp!]
\centering
\includegraphics[scale=1]{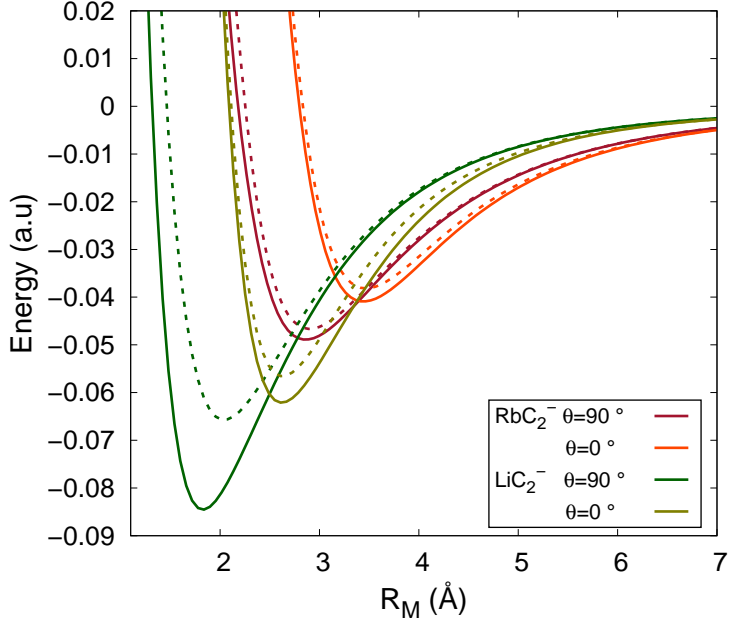}
\caption{Potential energy curves for the $^{1}A'$ (solid) and $^{3}A'$ (dashed) states of LiC$_{2}^{-}$ and RbC$_{2}^{-}$ for $\theta=0\degree$ and $90\degree$. The calculations were performed at the SA-CASSCF/MRCI+Q level of theory with the 1$s_{\textrm{\textrm{C}}}$ orbitals correlated while the distance $r$ was fixed at 1.27 $\angstrom$.}
\label{PECs}
\end{figure}
The potential energy curves for M-C$_{2}^{-}$ (where M = Li or Rb) at $\theta=0\degree$ and $\theta=90\degree$ are shown in Figure \ref{PECs} for both the singlet and the triplet $A'$ states and with the 1$s_{\textrm{C}}$ electrons correlated. The energy is given relative to the C$_{2}^{-} (^{2}\Sigma^{+}_{g})$ + M$(^{2}S)$ dissociation limit.
As can be seen, the triplet state is an excited state of the MC$_{2}^{-}$ molecular species. The adiabatic excitation energy at $\theta=0\degree$ is around 0.08 eV and 0.15 eV for Rb-C$_{2}^{-}$ and Li-C$_{2}^{-}$, respectively. This energy difference becomes 0.06 eV and 0.51 eV at $\theta=90\degree$, respectively. The depth of the potential well for the singlet and triplet states is largest at $\theta=90\degree$ and smallest at $\theta=0\degree$ for both systems. The equilibrium geometry of the MC$_{2}^{-}$ species in both spin multiplicities therefore corresponds to a T-shaped structure while the collinear structure ($\theta=0\degree$) corresponds to a saddle point. The larger difference between the singlet and triplet state for LiC$_{2}^{-}$ may be explained by the stronger mixing of the $2p_{C}$ and $ns_{Li}$ atomic orbitals, hence leading to larger energy gap between the $a'$ orbitals of LiC$_{2}^{-}$ and a larger excitation energy.
The value of the well depth for Li-C$_2^-$ and Rb-C$_2^-$ at the angles $\theta=0 \degree$ and $\theta=90 \degree$ in both spin multiplicities and with or without correlating the 1$s_{\textrm{C}}$ electrons is given in Table \ref{prop}. The effect of the 1$s_{\textrm{C}}$ correlation is rather small (less than 3\%) but depends on both the distance $R_{\textrm{M}}$ and on the angle $\theta$. It is worth noting that including the 1$s_{\textrm{C}}$ electrons into the correlation treatment leads to considerably larger computational effort: the number of contracted CSFs and the memory used increases by a factor of 5, and the calculation time by a factor of 14. However, correlating the core orbitals of $C_{2}^{-}$ was necessary to achieve convergence and avoid root flipping problems in the repulsive region of the PES.

The well depth of RbC$_{2}^{-}$ (1.33 eV) is similar to other systems such as Rb--CN$^{-}$ (1.20 eV), Rb-NCO$^{-}$ (1.25 eV), Rb--C$_{2}$H$^{-}$ (1.36 eV) and Rb--C$_{4}$H$^{-}$ (1.17 eV) \cite{Tomza2017}, but smaller than for Rb-OH$^{-}$ for which the interaction energy at the minimum is about 2 eV \cite{Gonzalez-Sanchez2008}. The well depth of LiC$_{2}^{-}$ is about 1.7 times larger than for RbC$_{2}^{-}$, in agreement with the trend observed for several other systems where Li compounds are found to be more strongly bound \cite{Kas2017,Tomza2017}. Unfortunately, no experimental results are available at the moment for comparison. It should be noted, however, that properties of C$_{2}^{-}$ such as the electron affinity, the equilibrium bond length and the excited states energies are all well described by the MRCI method \cite{Shi2016a}.

The anisotropy of the potential plays an important role in rotationally inelastic collisions of C$_{2}^{-}$ with Rb or Li. A rough estimation of the anisotropy may be obtained by taking the energy difference between extremum structures ($\theta=0 \degree$ and $\theta=90 \degree$). The values obtained are given in Table \ref{prop}. For RbC$_{2}^{-}$ ($X^{1}A'$) and RbC$_{2}^{-}$ (1$^{3}A'$) the $\Delta E_{0\degree-90\degree}$ values are similar and can be compared to other systems such as RbCN$^{-}$ (0.07 eV) and RbNCO$^{-}$ (0.14 eV), and RbC$_{2}$H$^{-}$ (1.01 eV), RbC$_{4}$H$^{-}$ (1.05 eV) \cite{Tomza2017} and RbOH$^{-}$ (1.20 eV) \cite{Gonzalez-Sanchez2008}. For singlet LiC$_{2}^{-}$ the $\Delta E_{0\degree-90\degree}$ is larger than for RbC$_{2}^{-}$ while being similar for the triplet state. Differences between the singlet and triplet PESs are therefore significantly larger for LiC$_{2}^{-}$. A more rigorous picture of the anisotropy can be developed by performing an expansion of the PES in Legendre polynomials, as will be described in section \ref{sec:CC}.

The long range interaction energy is dominated by the classical charge-induced dipole interaction, $V(R)=-\alpha/2R^{4}$. The polarisabilities of Rb and Li obtained by fitting this expression to our Rb-C$_{2}^{-}$ and Li-C$_{2}^{-}$ potentials at long range (from $R_{M}=18 \angstrom$ to $R_{M}=100 \angstrom$) are 165 $a_0^3$ for Li and 314 $a_0^3$ for Rb, in excellent agreement with the experimental values\cite{Molof1974} of 164 $a_0^3$ and 319 $a_0^3$, respectively. The agreement slightly decreases when the $1s_{\textrm{C}}$ correlation is neglected (168 $a_0^3$ and 308 $a_0^3$, respectively).

%
\begin{table}[h!]
\setlength{\tabcolsep}{6pt}
\def\arraystretch{1.05}
\centering
\begin{tabular}{cc|cc|cc|cc}
 & &  \multicolumn{2}{c|}{$E_{90\degree}^{min}$(eV)} & \multicolumn{2}{c|}{$E_{0\degree}^{min}$(eV)} & \multicolumn{2}{c}{$\Delta E_{0\degree-90\degree}$ (eV)}   \\
 & & 1$s_{\textrm{C}}^{closed}$ & 1$s_{\textrm{C}}^{open}$ & 1$s_{\textrm{C}}^{closed}$ & 1$s_{\textrm{C}}^{open}$ & 1$s_{\textrm{C}}^{closed}$ & 1$s_{\textrm{C}}^{open}$   \\
\hline
LiC$_{2}^{-}$ & $^{1}A'$ & -2.363  & -2.300 & -1.705 & -1.690 & 0.658 & 0.610    \\
              & $^{3}A'$ & -1.787  & -1.788 & -1.532 & -1.539 & 0.254 & 0.248  \\ 
RbC$_{2}^{-}$ & $^{1}A'$ & -1.327  & -1.331 & -1.109 & -1.113 & 0.219 & 0.218 \\
              & $^{3}A'$ & -1.263  & -1.268 & -1.030 & -1.037 & 0.234 & 0.231  

\end{tabular}
\caption{MRCI+Q well depth ($E^{min}$) for $\theta=90\degree$ and $0\degree$, and energy difference between $\theta=0\degree$ and $\theta=90\degree$ at $R_{min}$ ($\Delta E_{0\degree-90\degree}$). Values given with and without correlating the 1$s_{\textrm{C}}$ orbitals.}
\label{prop}
\end{table}

\section{Reactive collisions}
\label{Excited}
\begin{figure}[htbp!]
\centering
\begin{subfigure}[t]{0.4\textwidth}
\includegraphics[scale=0.7]{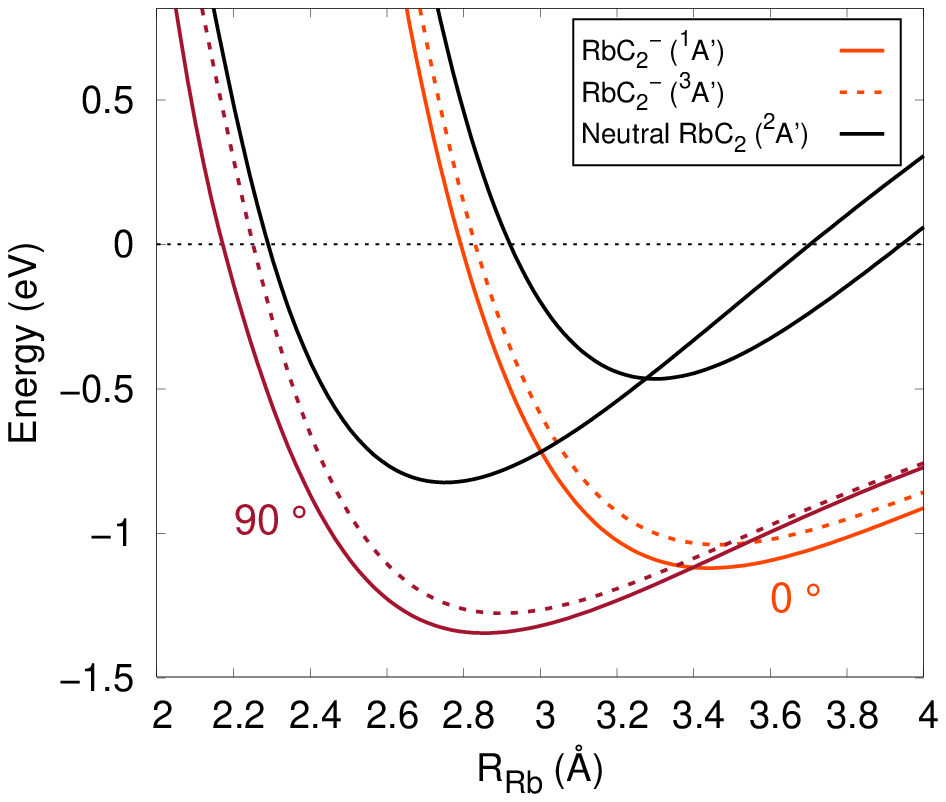}
\end{subfigure}%
\hspace{1cm}
\begin{subfigure}[t]{0.4\textwidth}
\includegraphics[scale=0.7]{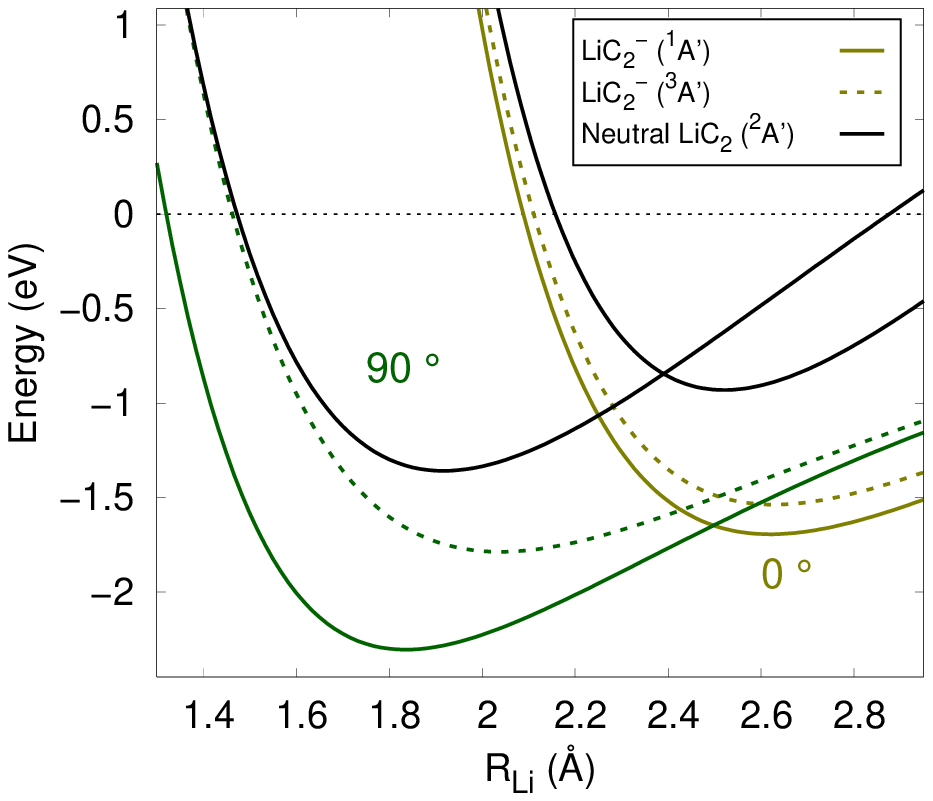}
\end{subfigure}
\caption{PECs of the anion singlet (thin solid line) and triplet (dashed line) states, and of the neutral doublet state (bold solid line) for $\theta=0 \degree$ and $90 \degree$ for the Rb--C$_{2}^{-}$ (left panel) and Li--C$_{2}^{-}$ (right panel) molecular systems. See text for computational details).}
\label{AetN}
\end{figure}

The only open reactive channel in the considered temperature range ($T<300$~K) is the associative electronic detachment (AED) reaction M($^{2}S$) + C$_{2}^{-}(^{2}\Sigma^{+}_{g})\rightarrow$ MC$_{2}(^{2}A_{1})$ + $e^{-}$. The reaction usually occurs when the PESs of the anion and neutral species cross, defining the autodetachment region. In this region, the spontaneous ejection of the excess electron usually takes place with a large rate\cite{Simons1998} (around 10$^{14}-10^{15}$ s$^{-1}$). In order to verify if the anion and neutral PESs cross, we have computed the potential energy surface for both neutral species, Rb-C$_2$ and Li-C$_2$. The CASSCF/MRCI+Q method with all electrons correlated was used for the calculations. The neutral ground state ($^{2}A'$) was incorporated along with the singlet and triplet anionic states into the state-averaged procedure in order to obtain a common CASSCF wave function. Again, the states are labelled according to the C$_{s}$ point group in which the calculations have been performed. The results are shown in Figure \ref{AetN} for the two extremum structures, namely $\theta=0\degree$ and $90\degree$.
As can be seen, the anionic and neutral curves cross above the dissociation limit of the entrance channel, suggesting a small rate for the AED reaction at room temperature and below. In particular, the rate should be considerably smaller than for OH$^{-}$ + Rb for which the PESs of the neutral and anion cross below the energy threshold for a large portion of the angular space \cite{Kas2016}. In the absence of energetically accessible crossings between anion and neutral PES, the electronic detachment can only occur through non-adiabatic couplings. The resulting detachment rate are usually small \cite{Simons1998} but depend on the reduced mass, the nature of the anionic wave function and the energy difference between anion and neutral PES \cite{Simonsl1981}. Therefore, differences between systems (Rb-C$_{2}^{-}$ and Li-C$_{2}^{-}$) and between singlet and triplet reactivity could be seen. \\   

\begin{figure}[htbp!]
\centering
\begin{subfigure}[t]{0.4\textwidth}
\includegraphics[scale=0.7]{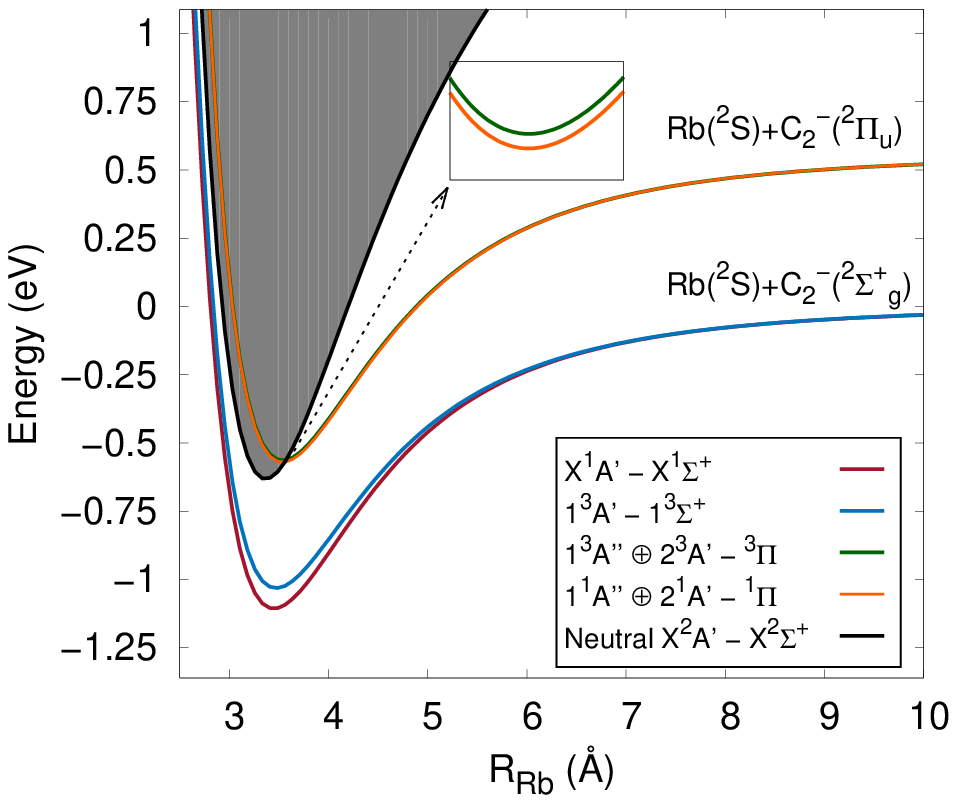}
\end{subfigure}%
\hspace{1cm}
\begin{subfigure}[t]{0.4\textwidth}
\includegraphics[scale=0.7]{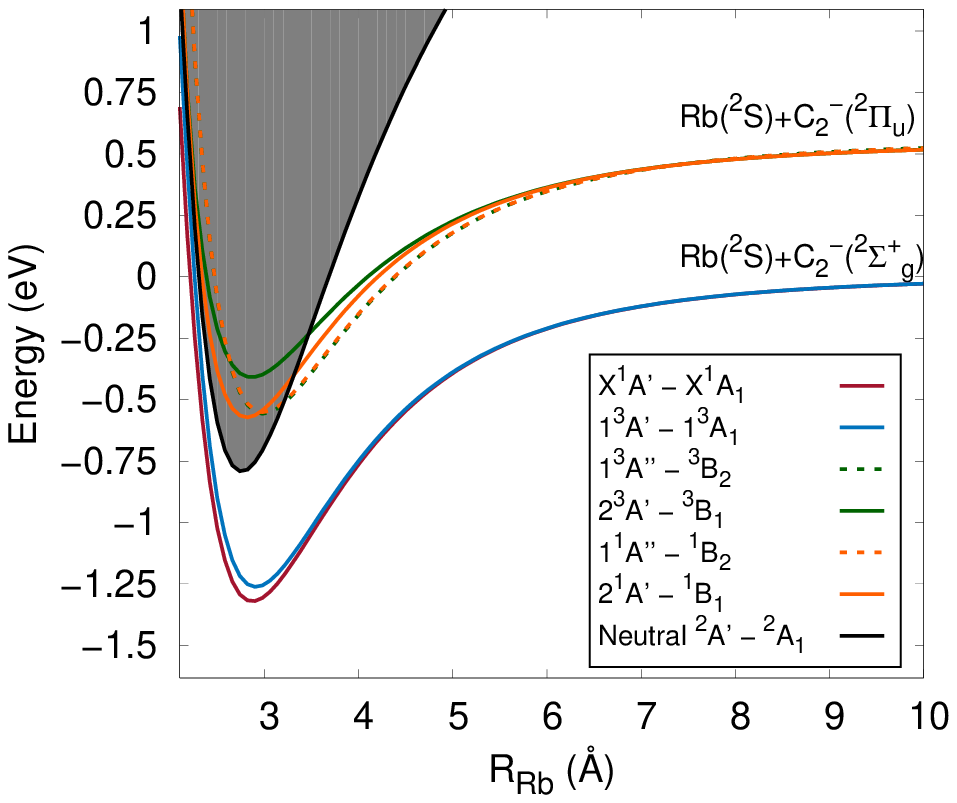}
\end{subfigure} \\
\begin{subfigure}[t]{0.4\textwidth}
\includegraphics[scale=0.7]{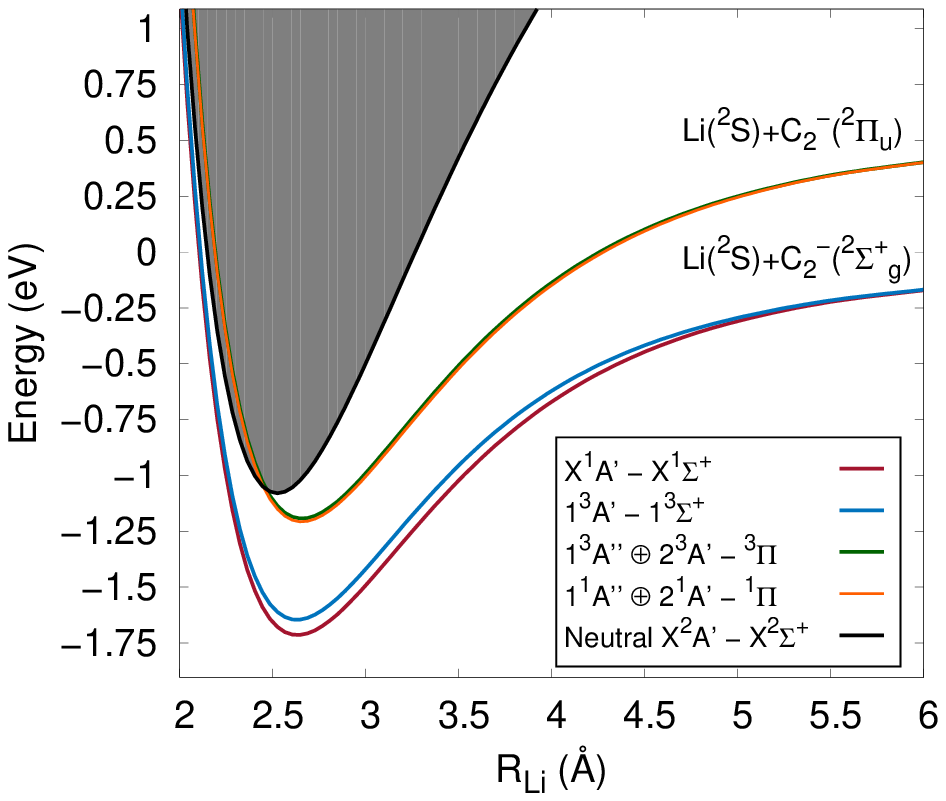}
\end{subfigure}%
\hspace{1cm}
\begin{subfigure}[t]{0.4\textwidth}
\includegraphics[scale=0.7]{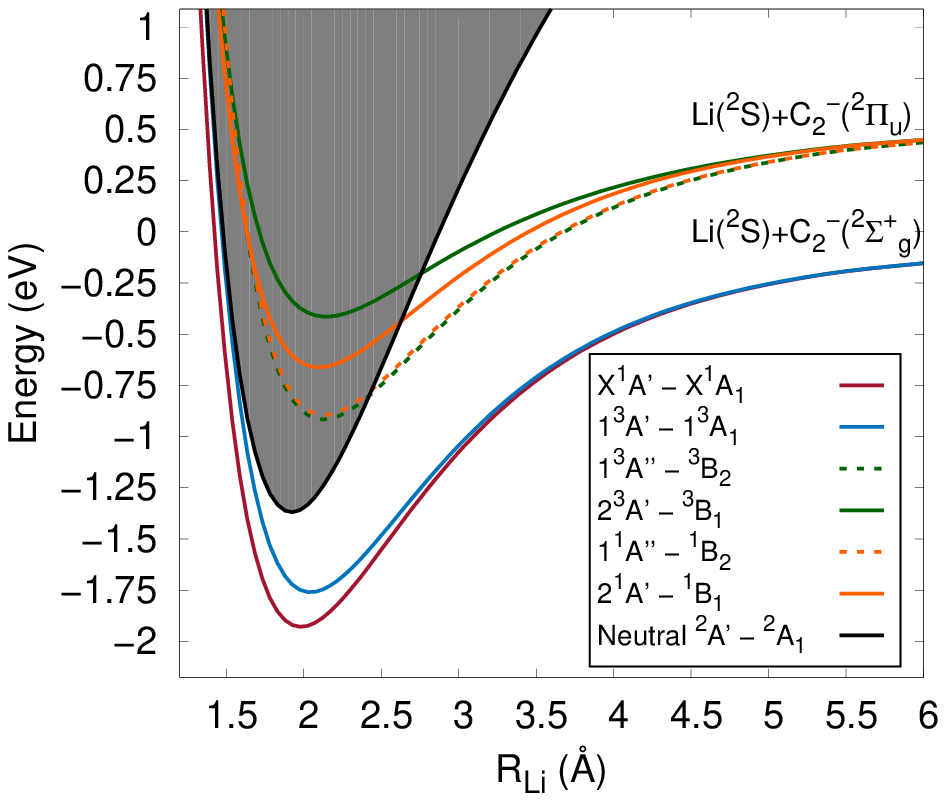}
\end{subfigure} \\
\caption{Low lying potential energy curves for the Rb--C$_{2}^{-}$ (top) and Li--C$_{2}^{-}$ (bottom) molecular species at $\theta=0\degree$ (left) and $90\degree$ (right). The two dissociation limits at which the molecular states correlate are shown. The grey region corresponds to the autodetachment region. Energy given relative to the ground state dissociation energy.}
\label{Ex}
\end{figure}

\begin{figure}[htbp!]
\centering
\includegraphics[scale=0.5]{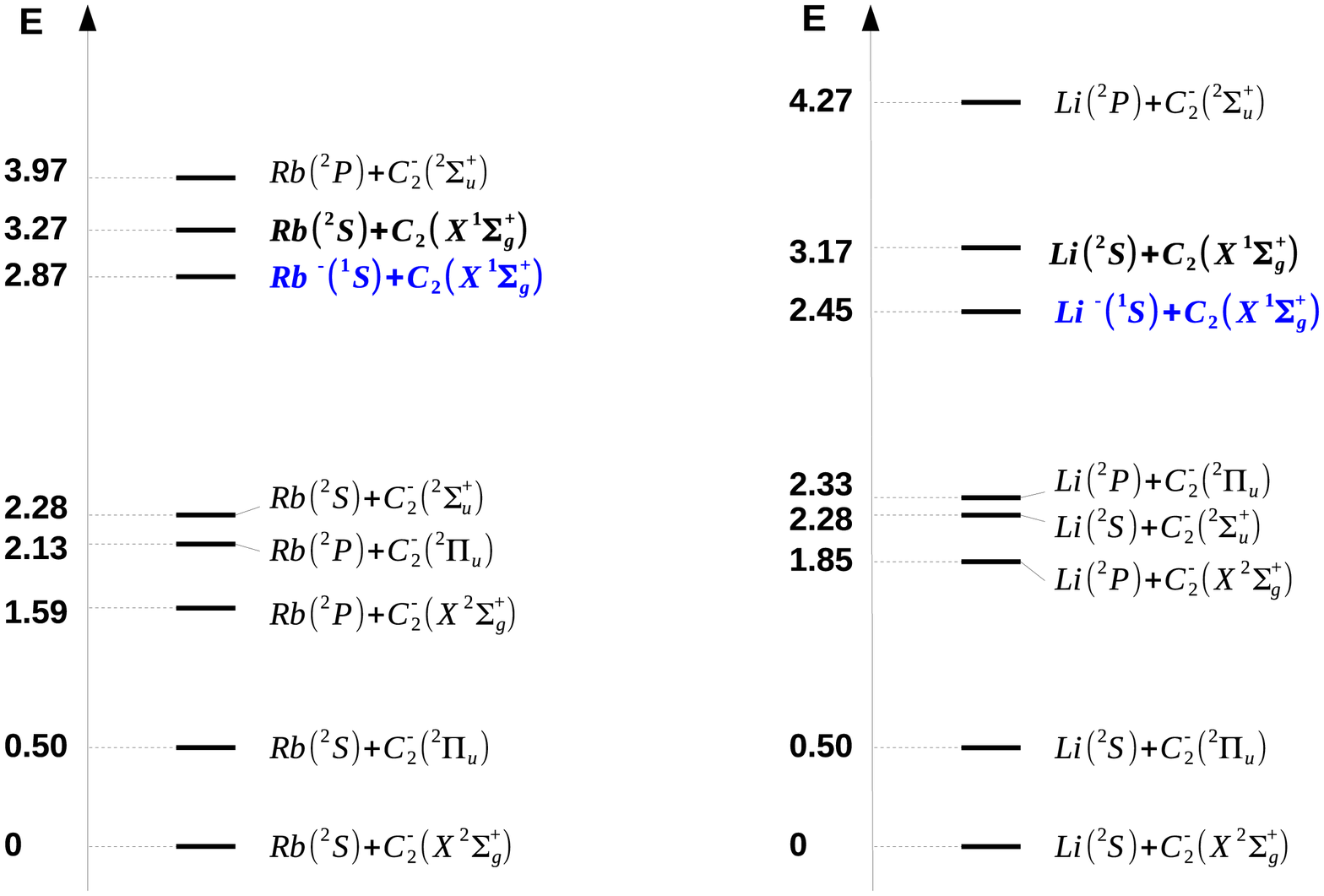}
\caption{Energy (in eV) of the different channels involving collision between Rb (left) or Li (right) with C$_{2}^{-}$. The neutral and charge transfer channels are shown in bold and blue, respectively. The different excitation energies and electron affinity of Li, Rb, and C$_{2}^{-}$ are taken from the NIST database.}
\label{Ex_tab}
\end{figure}

The C$_{2}^{-}$ anion supports several bound excited electronic states, allowing the use of Doppler thermometry to measure its translational temperature. The latter method implies an excitation of the C$_{2}^{-}$ anion into one of its stable excited states. By doing so, collisions between excited C$_{2}^{-}$ and Rb may occur, leading to a different reactivity. In order to investigate the possible outcome of such collisions, we have calculated the excited PESs of the molecular states that correlate to the M($^{2}S$)+C$_{2}^{-}$($^{2}\Pi_{u}$) dissociation channel, where $^{2}\Pi_{u}$ is the first electronic excited states of C$_{2}^{-}$. The CASSCF/MRCI+Q method with closed $1s_{\textrm{C}}$ orbitals has been used. The results are shown in Figure \ref{Ex} for both M=Li and M=Rb at $\theta=0 \degree$ and $90 \degree$. The term symbol of the molecular states are given in their ``true'' point group (i.e C$_{2v}$ and C$_{\infty v}$ for $\theta=90 \degree$ and $0 \degree$, respectively) and, in parenthesis, in the C$_{s}$ point group which defines intermediate geometries. The grey region corresponds to the autodetachment region defined as $E> E_{\textrm{neutral}}$.          
As can be observed the excited curves cross the autodetachment region below the energy of the entrance channel. This suggests that the AED reaction M($^{2}S$) + C$_{2}^{-}(^{2}\Pi_{u}) \rightarrow$ MC$_{2}(^{2}A_{1})+e^{-}$ is likely to occur with a large probability. Furthermore, the rate should be close to the Langevin rate since all collisional angles lead to crossing into the autodetachment region. Higher excited states of C$_{2}^{-}$ such as the $^{2}\Sigma^{+}_{u}$ state may be used for Doppler thermometry instead of the first excited $^{2}\Pi_{u}$ state. However, since the molecular states arising from the M($^{2}S$)+C$_{2}^{-}$($^{2}\Sigma^{+}_{u}$) entrance channel lie above the one shown in Figure \ref{Ex}, they will ultimately cross the autodetachment region. The same can be said for collisions involving Rb($^{2}P$) for which the entrance channel lies between the two discussed above. The various entrance channels involving excited states are shown in Figure \ref{Ex_tab} for Rb and Li with their respective energy relative to the ground state channel.
Reactive channels involving bond breaking of C$_{2}^{-}$ will not be accessible since the dissociation energy is about 8.1 eV (obtained from AWCVQZ/CASSF/MRCI+Q calculations). Therefore, even the channel involving both excited M($^{2}P$) and C$_{2}^{-}(^{2}\Sigma^{+}_{u})$ states lies several eV below the MC + C$^{-}$ dissociation channel, which remain inaccessible in the energy regime considered here. However, the dissociative electronic detachment (DED) reaction M($^{2}P$) + C$_{2}^{-}(^{2}\Sigma^{+}_{u})\rightarrow$ M($^{2}S$) + C$_{2}$($^{1}\Sigma_{g}^{+}$) + $e^{-}$ and charge transfer reaction M($^{2}P$) + C$_{2}^{-}(^{2}\Sigma^{+}_{u})\rightarrow$ M$^{-}$($^{1}S$) + C$_{2}$($^{1}\Sigma_{g}^{+}$) become energetically accessible. Both reactions will compete with the AED channel, but the latter is likely to be dominant since the system will first enter into the autodetachment region before it can exit via the CT or DED channels. \\
In the case of Rb, the spin-orbit coupling may influence the different potential energy curves. In order to verify that the interaction between the $A'$ and $A''$ states does not allow the anionic states to drop below the neutral curve, we performed spin-orbit calculations using the Breit-Pauli Hamiltonian with the state interacting method for MRCI wavefunctions implemented in {\small MOLPRO} \cite{Berning2000}. Our calculations show that the resulting SO interaction is negligible and does not affect the dynamics in the sense that the AD region is still reached.

In conclusion, collisions involving excited species should predominantly lead to AED reactions. Experimental measurements could verify these hypotheses. Distinguishing the DED from AED products is not straightforward since both reactions lead to losses, and a specific detection method would therefore be needed. Nonetheless, the CT products M$^{-}$ should be easier to detect. The AED rate from both $^{2}\Sigma^{+}_{u}$ and $^{2}\Pi_{u}$ entrance channels should be similar and close to the Langevin rate since no additional long range interactions have to be taken into account. As has already been observed for N$_{2}^{+}$ \cite{Hall2012} and OH$^{-}$ \cite{Kas2016}, the rate should therefore be larger for the channels involving M($^{2}P$).

\FloatBarrier
\section{\label{sec:CC}Inelastic collisions}
\label{Dyn}
\begin{figure}[htbp!]
\centering
\includegraphics[scale=1]{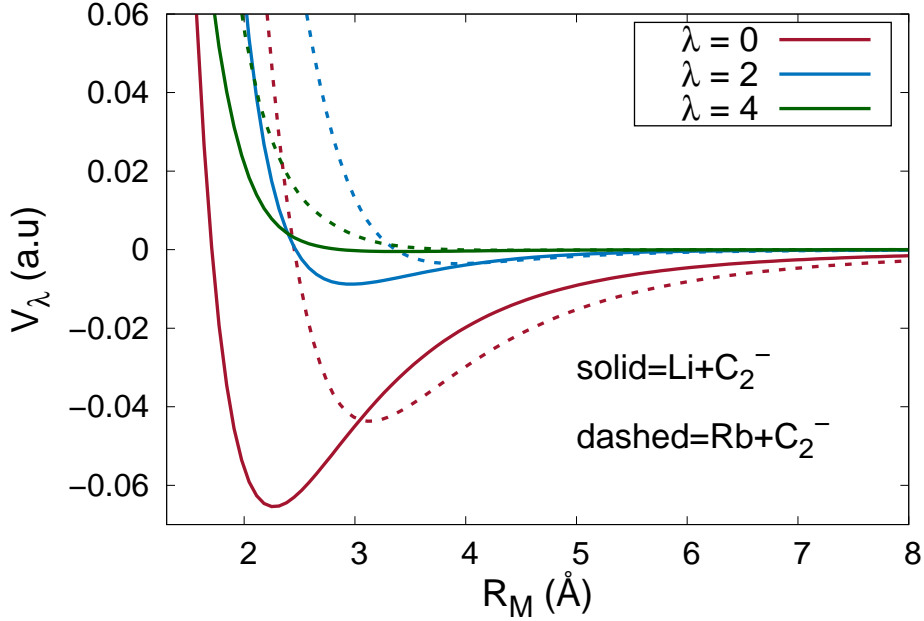}
\caption{Legendre coefficients $v_{\lambda}(R_M)$ for the singlet PES of Rb--C$_{2}^{-}$ (dashed) and Li--C$_{2}^{-}$ (solid). Only the first 3 terms of the expansion are shown.}
\label{Leg}
\end{figure}

Since collisions of ground state C$_2^-$ with either Li or Rb atoms are non-reactive at low temperatures, we next investigate the possibility of cooling the rotational degree of freedom of C$_2^-$ through inelastic collisions. Rotationally inelastic cross sections were calculated by means of the close-coupling (CC) approach \cite{Arthurs1960}. The method has been widely used on a variety of molecular systems and details on the underlying theory may be found elsewhere \cite{Flower2007}. Here, we only briefly recall the main equations of the CC approach. The method is based on an expansion of the total wave function into an angular basis set:
\begin{equation}
\Psi=R_{\textrm{M}}^{-1} \sum_{\alpha}|\alpha \rangle \chi_{\alpha}(R_{M})
\end{equation}
where R$_{M}$ is the radial distance and $\chi_{\alpha}$ are radial functions describing the nuclear motion. In the case of atom-diatom (rigid rotor) collisions, the angular functions $|\alpha \rangle$ are described by the quantum numbers $l,j,J$ which are eigenvalues of the angular momentum operator of the colliding system, the angular momentum operator of the rigid rotor and the total angular momentum operator, respectively. We thus have: $|\alpha\rangle=|ljJ\rangle$. Integration of the time-independent Schr\"odinger equation over the angular variables leads to a set of coupled differential equations:
\begin{equation}
-\frac{\hbar^{2}}{2\mu}\frac{d^{2}\chi_{ljJ}}{dR_{M}^{2}}+\sum_{l'j'}\langle l'j'J|\frac{\textbf{j}^{2}}{2I}+V(R_{M},\theta)+\frac{\textbf{L}^{2}}{2\mu R_{M}^{2}}|ljJ\rangle \chi_{ljJ}=E\chi_{ljJ}
\end{equation} 
where the first term is the kinetic energy operator, $E$ is the total energy, $\textbf{j}^{2}/2I$ is the angular momentum operator of the rigid rotor, $V$ is the potential energy surface and $\textbf{L}$ is the angular momentum describing the relative motion of the colliding partners. 
The potential $V(R_{M},\theta)$ can be expanded over Legendre polynomials:
\begin{equation}
V(R_{M},\theta)=\sum_{\lambda}v_{\lambda}(R_M)P_{\lambda}(\cos\theta).
\end{equation}        
The three largest coefficients $v_{\lambda}(R_M)$ obtained for the Rb--C$_{2}^{-}$ and Li--C$_{2}^{-}$ singlet potentials are shown in Figure \ref{Leg}. 

The potential allows for the coupling between different $j$ levels through the following matrix elements:
\begin{equation}
\langle ljJ| V(R_{M},\theta)|l'j'J\rangle
\end{equation}
the $\lambda$, $j$ and $j'$ values are subjected to the triangular inequalities $|j-j'|\leq \lambda\leq j+j'$ \cite{Flower2007}. 
Systems which exhibit strong anisotropy (large $v_{\lambda}$ for high $\lambda$ values) will have larger inelastic cross sections. Moreover, in the present case of atom-homonuclear diatom collision the potential $V(R,\theta)$ only contains even $\lambda$ terms (see Figure \ref{Leg}) so that only $\Delta j=2$ transitions are allowed. In addition, due to symmetry upon permutation of the total wavefunction, only rotational states with even values of $j$ are present in C$_{2}^{-}$ when considering the most abundant isotope $^{12}$C. \\
The rotational constant and higher order terms for C$_{2}^{-}$ are taken from Ref. \onlinecite{Herzberg1968}. These values have been used to construct the angular basis for the close-coupling calculations with inclusion of rotational states up to $j=30$.
%
%
The CC equations were solved by means of the hybrid modified log-derivative Airy propagator \cite{Manolopoulos1987} implemented in {\small MOLSCAT} \cite{Molscat} over a grid of distances from 3~$a_0$ to 200~$a_0$. For distances below 50~$\angstrom$, the usual log-derivative propagator is used while for R$_{M}> 50$ $\angstrom$ the Airy method is employed. 
Inelastic cross sections were obtained for collision energies up to 400 cm$^{-1}$ for Li--C$_2^-$ and Rb--C$_2^-$ in the singlet and triplet states. 

\begin{figure}[htbp!]
\centering
\hspace{-3cm}
\begin{subfigure}[t]{0.4\textwidth}
\includegraphics[scale=0.7]{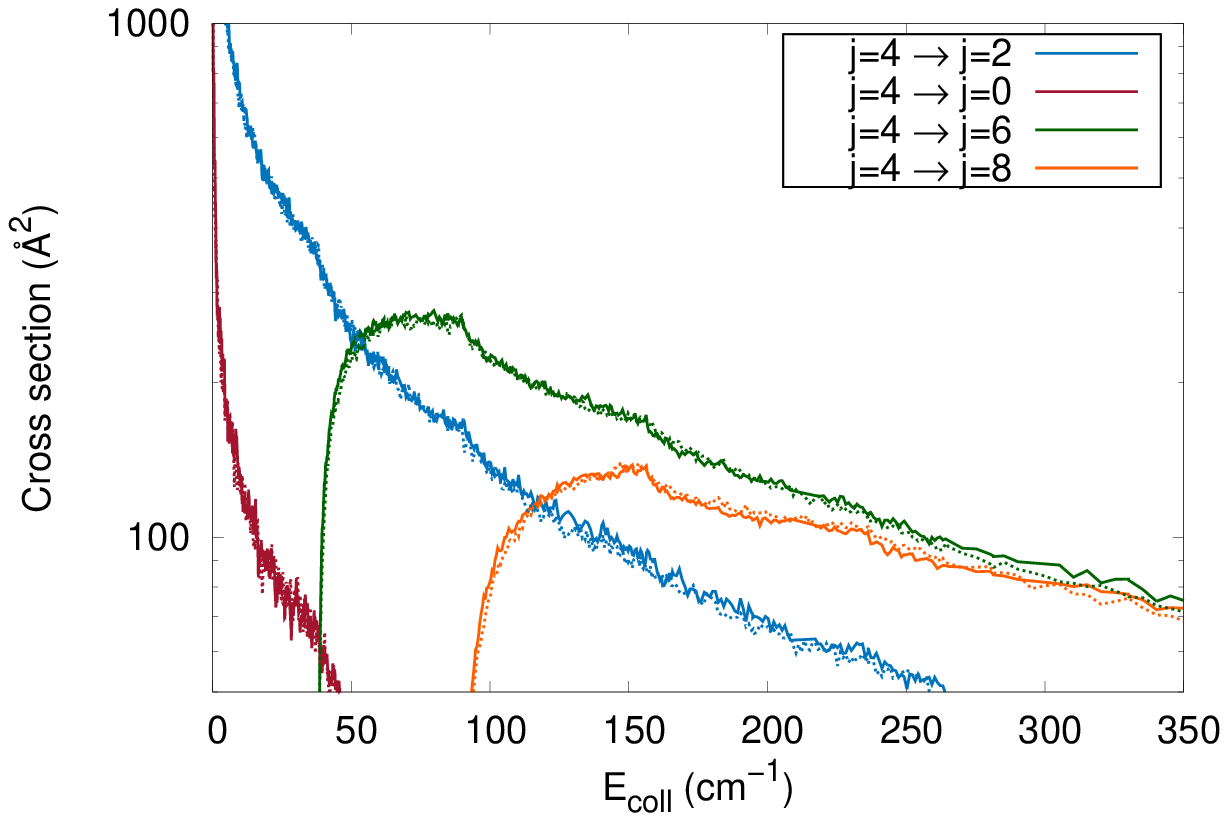}
\end{subfigure}%
\hspace{2cm}
\begin{subfigure}[t]{0.4\textwidth}
\includegraphics[scale=0.7]{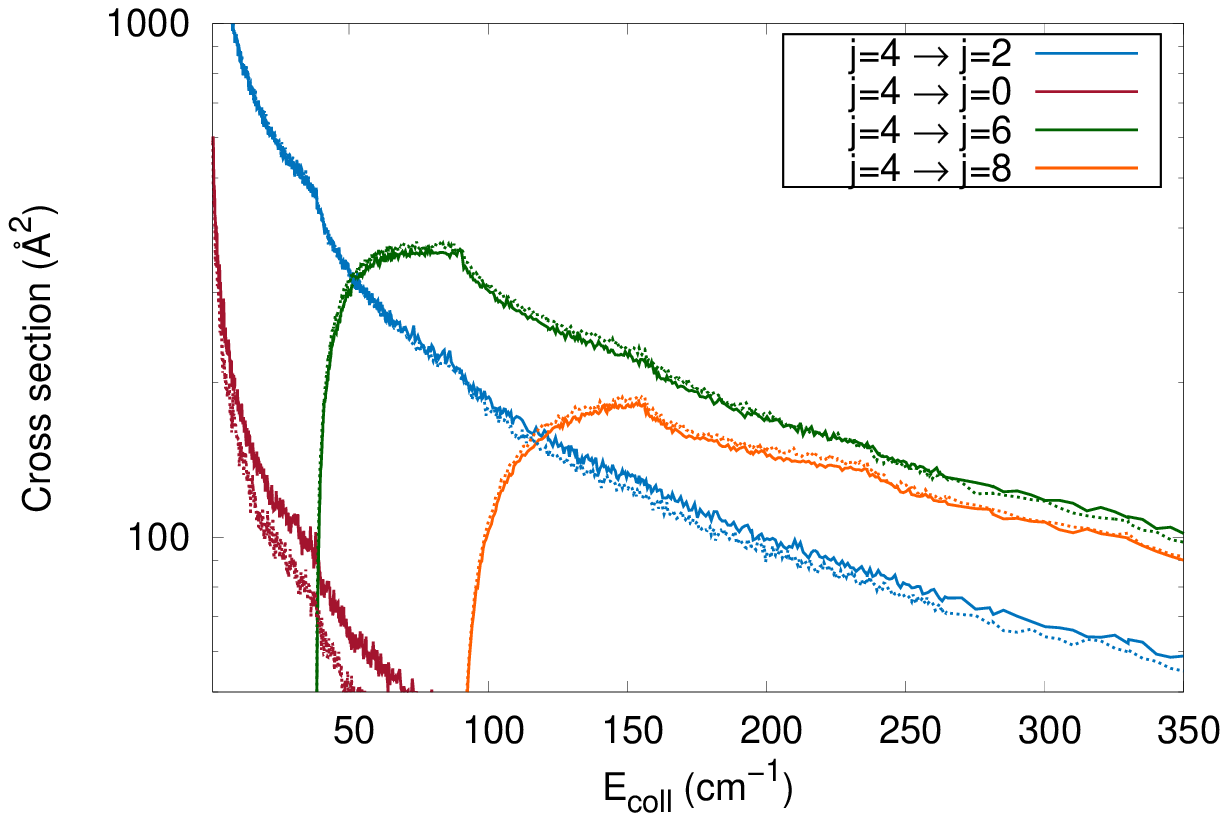}
\end{subfigure}
\caption{Rotationally inelastic cross sections for  Li--C$_{2}^{-}$ (left) and Rb--C$_{2}^{-}$ (right) scattering in the initial state $j=4$. Full lines, singlet state; dashed lines, triplet state.}
\label{fig_inel_cs_j4}
\end{figure}

Rotationally inelastic cross sections for scattering of C$_2^-$ with Li and Rb atoms are shown in Fig. \ref{fig_inel_cs_j4} for the initial state $j=4$, the most populated rotational state at $T=100$ K. 
As expected, the excitation cross sections to the states $j=6$ and $j=8$ show a sharp increase at the threshold, when the transition becomes energetically possible, and both excitation and de-excitation cross sections decrease when the difference $\Delta j$ between the initial and final states increase. The cross sections present a dense resonance structure, which are a combination of shape and Feshbach resonances. 

It can be seen that the cross sections for Li and Rb atoms are of similar magnitude although they are larger for Rb. For Li--C$_2^-$ collisions, the differences between the singlet and triplet PESs on the dynamics are negligible. On the other hand, larger differences occur for Rb--C$_2^-$. These differences depend both on the initial rotational state and on the $\Delta j$ between the initial and final rotational states. Overall, the differences between the cross sections for both spin multiplicities are surprisingly small given the variation in the PESs discussed above. This may be explained by the fact that at low collision energies the long range interaction dominates the dynamics. In what follows, the cross sections are averaged over the spin multiplicity using the statistical ratio 1:3.

\begin{figure}[htbp!]
\centering
\hspace{-3cm}
\begin{subfigure}[t]{0.4\textwidth}
\includegraphics[scale=0.7]{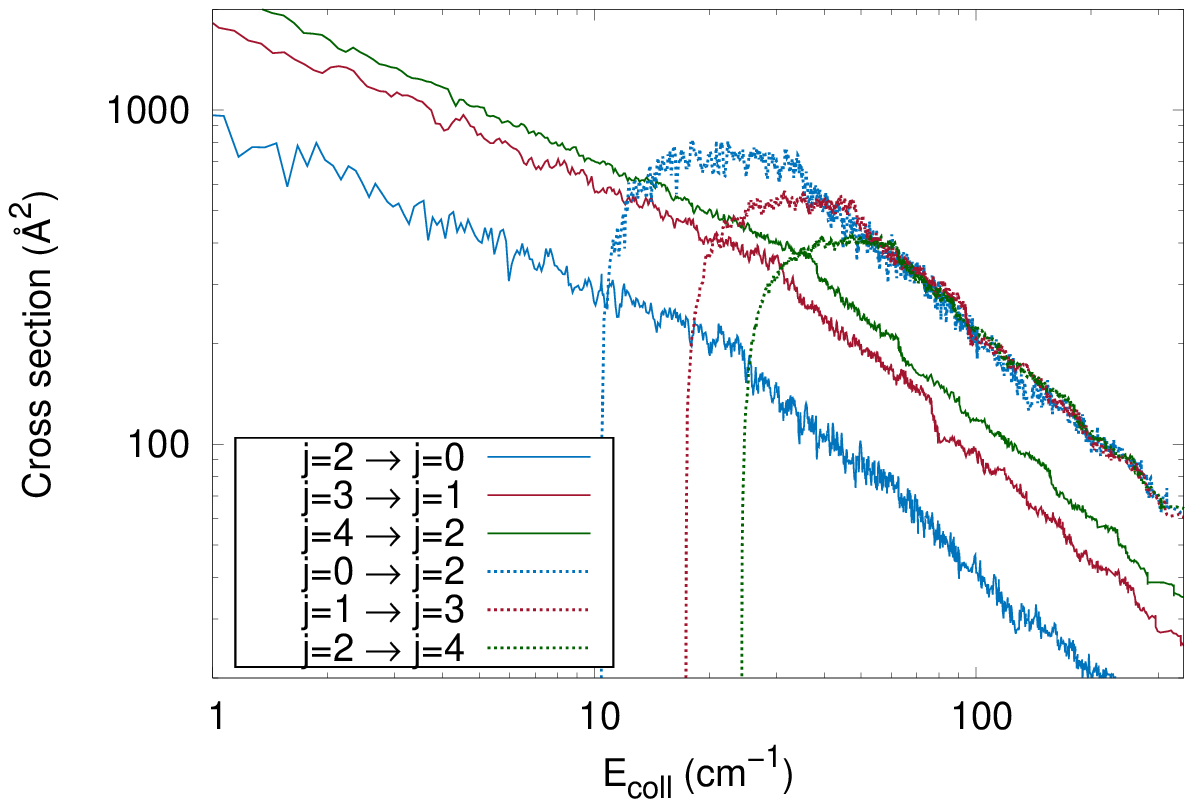}
\end{subfigure}%
\hspace{2cm}
\begin{subfigure}[t]{0.4\textwidth}
\includegraphics[scale=0.7]{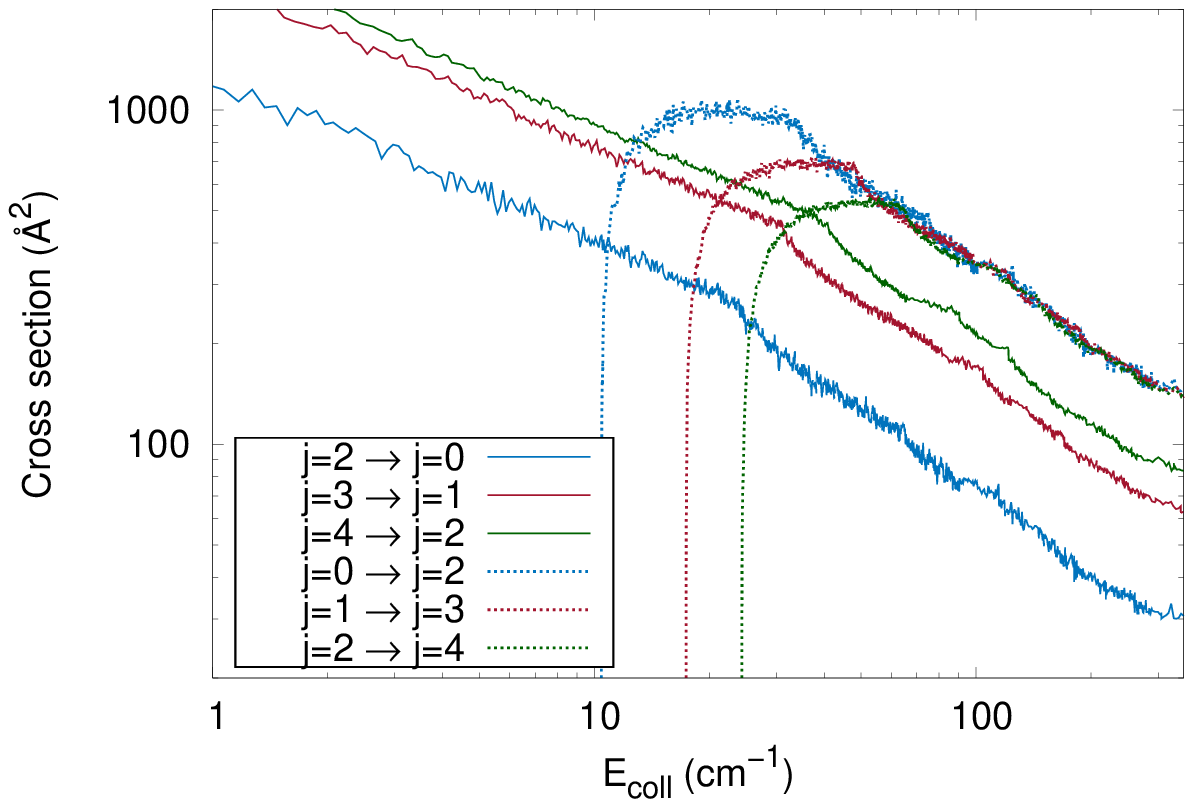}
\end{subfigure}
\caption{$\Delta j=2$ rotationally inelastic cross sections for  Li--C$_{2}^{-}$ (left) and Rb--C$_{2}^{-}$ (right) collisions in various initial states. Full lines, de-excitation; dashed lines, excitation.}
\label{fig_inel_cs_Deltaj2}
\end{figure}

Fig. \ref{fig_inel_cs_Deltaj2} shows the $\Delta j=2$ inelastic cross sections for various initial rotational states of C$_2^-$. We observe again that the cross sections are consistently larger for Rb--C$_2^-$ than for Li--C$_2^-$, although the shape of the cross sections is similar. In addition, while the excitation cross sections decrease with an increase in the initial rotational state $j_i$ of C$_2^-$, the de-excitation cross sections increase with $j_i$. The same behaviour was also observed for Rb-OH$^-$ collisions \cite{Gonzalez-Sanchez2015}. 

All the results presented here were obtained with the most accurate PESs, {\it i.e.} those for which the $1s_{\mathrm{C}}$ electrons are correlated. We performed additional calculations on the PESs without $1s_{\mathrm{C}}$ correlation and found that the effect on the cross sections is limited but can reach a few percent. 

\begin{figure}[htbp!]
\centering
\hspace{-3cm}
\begin{subfigure}[t]{0.4\textwidth}
\includegraphics[scale=0.7]{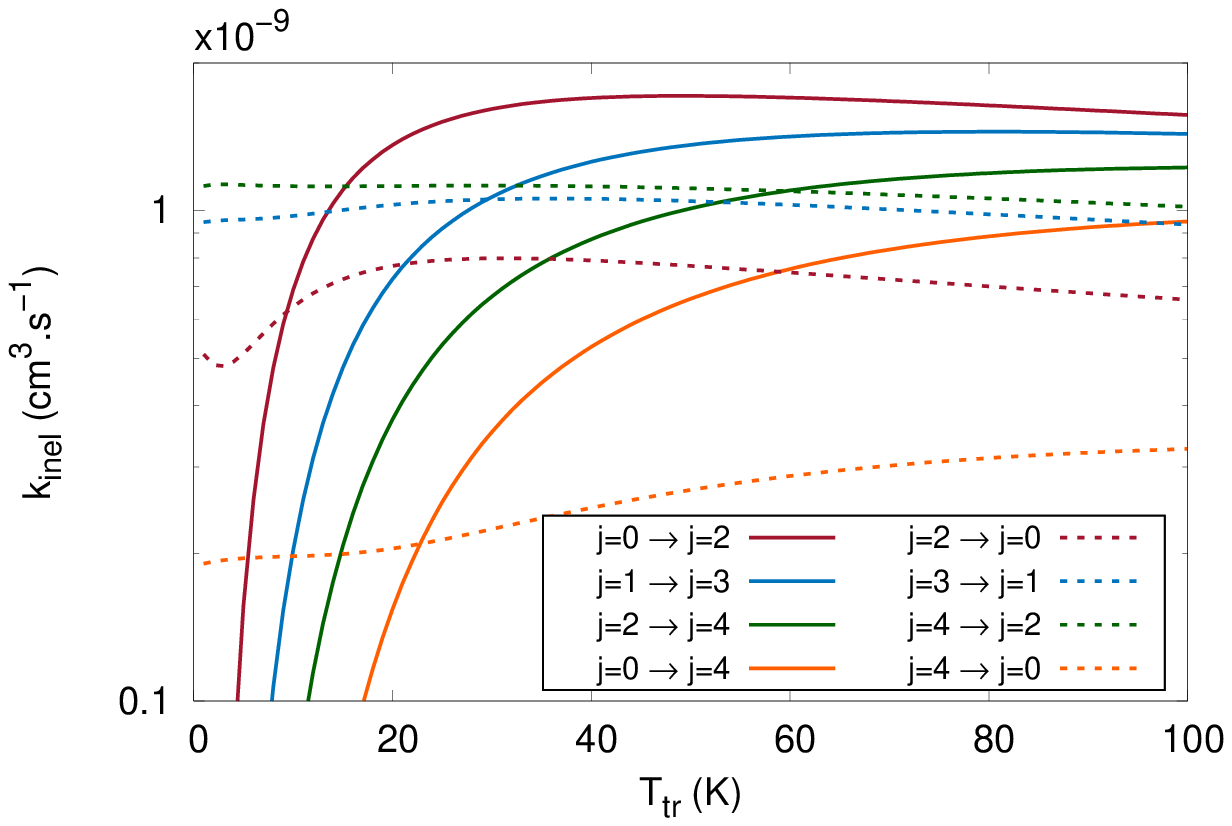}
\end{subfigure}%
\hspace{2cm}
\begin{subfigure}[t]{0.4\textwidth}
\includegraphics[scale=0.7]{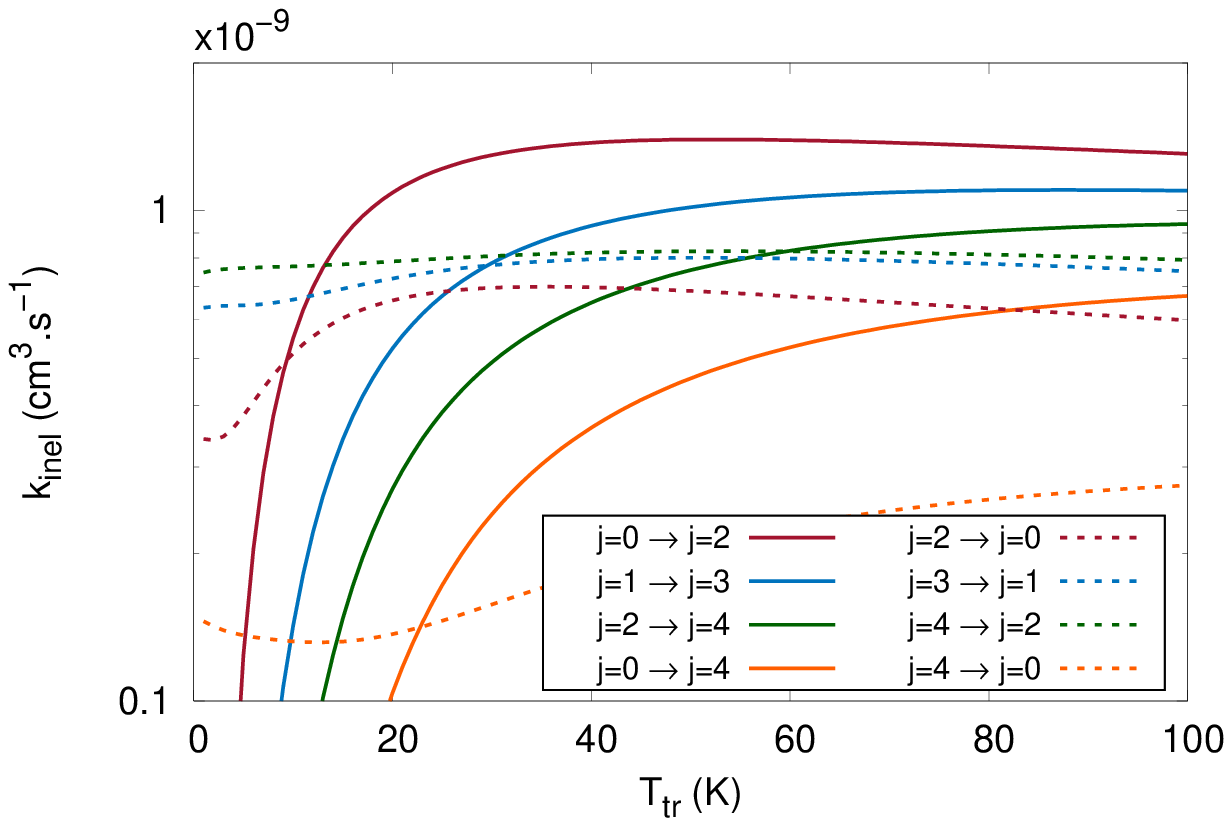}
\end{subfigure}
\caption{Rate coefficients for transitions involving the first four rotational levels of C$_2^-$ in collisions with Li (left) or Rb (right).}
\label{fig_rates}
\end{figure}

The cross sections were averaged over a Maxwell-Boltzmann energy distribution to calculate the state-to-state rate coefficients for temperatures between $1$ K and $100$ K:
\begin{equation}\label{eq_rate}
k(T) = \Big(\frac{2}{k_BT}\Big)^{3/2} \frac{1}{\sqrt{\pi\mu}} \int_0^\infty E e^{-E/k_BT} \sigma(E) \ dE \ .
\end{equation}
The rate coefficients for the transitions involving the first four rotational levels of C$_2^-$ are shown in Fig. \ref{fig_rates}.
The rate coefficients for excitation increase rapidly at low temperature, as expected from the cross sections shown in Figs. \ref{fig_inel_cs_j4} and \ref{fig_inel_cs_Deltaj2}. On the other hand, the de-excitation cross sections show little variation over the range of temperatures 1-100~K, with values around $10^{-9}$ cm$^3$s$^{-1}$. The de-excitation rates for transitions $\Delta j=4$ are approximately three times smaller than for $\Delta j=2$, except for the transition $j=2\rightarrow 0$ which is much smaller. It can also be observed that the rate coefficient for a given transition is larger for Li than for Rb. While the cross sections were larger for Rb than for Li, this effect is more than compensated by the presence of the reduced mass in Eq. (\ref{eq_rate}). These values can be compared to the Langevin rates of $5.01\times 10^{-9}$ cm$^3$s$^{-1}$ for Li-C$_{2}^{-}$ and $3.75\times 10^{-9}$ cm$^3$s$^{-1}$ for Rb-C$_{2}^{-}$.

Based on these large inelastic rate coefficients, the rotational cooling of C$_2^-$ with ultracold Li or Rb atoms can be expected to be efficient. The present results are also expected to be representative of the alkali series as the PESs for the interaction of C$_2^-$ with other alkali atoms (Na, K, Cs) are likely to be similar to those for Li--C$_2^-$ and Rb--C$_2^-$.

%
%

\FloatBarrier

\section{\label{sec:con}Conclusion}
We have carried out a theoretical study of cold Li--C$_2^-$ and Rb--C$_2^-$ collisions in the context of hybrid trap experiments. 
We have investigated the associative detachment reaction which, in the considered energy regime, is the only accessible reactive channel. Based on the potential energy surfaces of the ground singlet and triplet states of the anion and the corresponding neutral, we predict the rate of associative detachment reaction to be very small. In addition, we have shown that the reaction is likely to occur with a rate close to capture theory when considering collisions involving excited electronic states of C$_{2}^{-}$ and/or Rb or Li. By reducing the fraction of excited Rb in the MOT (decreasing the laser intensity or using a DarkSPOT) or by using an alternative trap ($e.g.$ a dipole trap) one could either decrease or hinder the loss through the excited channels. The associative detachment reaction will also be important when considering collisions with excited electronic states of C$_{2}^{-}$ which could be used for fluorescence imaging or Doppler thermometry. The atom trap should thus be switched off before performing such measurements in order to avoid losses. Alternatively, the electronic state dependence of the reactivity could be investigated. Furthermore, we have calculated the rotationally inelastic cross sections for scattering of C$_{2}^{-}$ with Li and Rb atoms. Similarly to the Rb--OH$^{-}$ case, we found the cross sections and corresponding rate constants to be rather large, suggesting a fast thermalisation of the ions rotational temperature considering low translational temperatures. The cross sections obtained when using either the singlet and triplet potential are surprisingly similar. When considering the most abundant $^{12}$C isotope, only even $j$ rotational states are present in C$_{2}^{-}$ and only $\Delta j=$ even transitions are allowed. Therefore, producing a sample of anions in their ground rotational state should be feasible if the translational temperature is sufficiently low. In addition, since C$_{2}^{-}$ is a homonuclear diatomic species, light-induced rotational dipole transitions are forbidden, meaning that excitation of the rotational state through black-body radiation should be hindered, or at least very small. The knowledge of the scattering cross sections  and the typical trap dynamics (which leads to non-thermal energy distributions \cite{Holtkemeier2017}) could be used to build accurate models of the collision dynamics occurring in hybrid trap experiments. Finally, based on similarity of the potential for different anions interacting with various alkali atoms\cite{Kas2017,Tomza2017} our conclusions are expected to hold for collisions of C$_2^-$ with Na, K or Cs atoms.

\acknowledgments
The Fonds National de la Recherche Scientifique de Belgique (FRS-FNRS) is greatly acknowledged for financial support (FRIA grant and IISN 4.4504.10 project). 
Computational resources have been provided by the Shared ICT Services Centre, Universit\'e libre de Bruxelles, and by the Consortium
des \'Equipements de Calcul Intensif (C\'ECI), funded by the Fonds de la
Recherche Scientifique de Belgique (F.R.S.-FNRS) under Grant No.
2.5020.11.

\bibliographystyle{apsrev4-1}

%

\end{document}